\documentclass[twocolumn,prl,aps,superbib,tightenlines,floatfix]{revtex4}
\usepackage{amsfonts,amsmath,amssymb,amsthm,bm}
\usepackage{graphicx}
\usepackage[dvipsnames,usenames]{color}
\begin{document}
\bibliographystyle{apsrev}
\title{Current Streamline Flow on Current-induced Effects in Highly Asymmetric Molecular Junctions.}
\author{Bailey~C. Hsu}
\author{Allen Tseng}
\author{Yu-Chang Chen}
\email{yuchangchen@mail.nctu.edu.tw}
\affiliation{Department of Electrophysics, National Chiao Tung University, 1001 University Road,
Hsinchu 30010, Taiwan}
\begin{abstract}

From first-principles approaches, we illustrate that the current-induced
forces and the selection rule for inelastic effects are highly relevant
to the current density in an asymmetric molecular junction.
The curved flow of current streamline around the asymmetric
molecule may induce a net torque, which tends to rotate the benzene
molecule, similar to the way a stream of water rotates a waterwheel.
Thus, the Pt/benzene junction offers a practical system in the exploration of
the possibility of atomic-scale motors. We also enumerate examples to
show that the use of selection rule can lead to misjudgement of the
importance of normal modes in the inelastic
profiles when the detailed information about the current density
is not considered.

\end{abstract}
\maketitle

Electron transport in a single-molecule molecular junction, where the molecule
is sandwiched between two electrodes, has been investigated extensively
in the pursuit of extreme device miniaturization~\cite{Tao,Nitzan,Majumdar1,Diventrarev}.
A major concern for single-molecule junctions is the fundamental properties
of current-induced effects due to nonequilibrium electron transport
at finite biases. These effects are efficient tools to explore the single-molecule
signatures from quantum mechanical perspectives, and thus are important
and interesting from theoretical and experimental points of view.
For example, the question on whether the current-induced forces on atoms are
conserved has been raised~\cite{Diventraforce}. Through molecular dynamics
simulations, it has been recently reported that the current-induced
forces are generally not conserved~\cite{Brandbyge,Todorov}. The authors
show that it is possible to construct atomic-scale systems where the
current-induced forces can be used to rotate the atoms. Therefore,
the current-induced forces are not conserved due to the nonzero
net work done by the current-induced forces~\cite{Todorov}.

The current-induced effects with nuclear degrees of freedom of molecules
can enrich the understanding of the interactions between electrons
and molecular vibrations. In the hydrogen junctions, the mode identification
and the selection rule for modes that significantly contribute to IETS are
related to the large component of vibration along the direction of electron
transport ($z$-direction)~\cite{chenH2}. Due to the short length of the junction,
electrons are transported quasi-ballistically in the molecular
junction~\cite{Galperin}. Consequently, both the interaction between electrons
and local ions (electron-vibration interaction) and the induced local heating
need to be considered.~\cite{Chen,Huang,Kushmerick,Sergueev}.
Electron-vibration interaction reveals the junction signatures
through the inelastic electron tunneling spectroscopy (IETS)
~\cite{Kushmerick1,Tsutsui}, while local temperature
resulting from local heating can be obtained by equilibrating
the heat generation in the junction with heat dissipation to the electrodes.
Recent report shows that local heating leads to a stronger suppression of the
current at longitudinal vibrational modes for a 4-Al atomic wire~\cite{Hsu}.

Moreover, Kiguchi~\textit{et~al.} have measured a high conductance in a single-molecule
junction where the benzene molecule is connected directly to the platinum
electrodes (Pt/benzene junction)~\cite{Kiguchi}. This demonstrates the feasibility
of a carbon-metal link. The authors have concluded that the benzene molecule forms a direct bond
with the electrodes. This is based on the analysis of the conductance histogram
and IETS, where the vibrational modes in the inelastic profiles have been verified by
isotope substitution. The relaxed configuration of the Pt/benzene junction, as
we shall show later, loses mirror symmetry. The highly tilted benzene molecule
causes the streamline flow of the current to curve considerably to one side
of the benzene ring. This could cause the unbalanced current-induced forces,
which tend to rotate the benzene molecule. Moreover, the mode selection rule
for IETS in the highly tilted benzene molecule junction becomes intricate.
The selection rule based on the component of vibrations along the z-direction
turns out to be inappropriate. Determining the importance the contribution of normal modes
to IETS based only on the $z$-component of vibrations will lead to misjudgement.
This suggests the necessity of looking into the detailed current density,
especially in highly asymmetric molecular junctions, such as the Pt/benzene
junction.

In this Letter, we demonstrate that the current-induced effects
are highly relevant to the details of the current density.
This is particularly important in the Pt/benzene junction because of its
highly asymmetric configuration. From first-principles approaches,
we investigate the impact of asymmetric current streamline flow on
the current-induced forces and the selection rule for IETS in the
Pt/benzene junction. It should be noted that our
theoretical method has an advantage over the DFT+nonequilibrium
Green function method since the latter does not
offer information about \emph{current density}.
We show that the curved current streamline flow
around the molecule may induce a net torque, which tends
to rotate the benzene molecule. To compare with the experimental IETS data,
we calculate the inelastic profile of conductance $(dI/dV)$ and the derivative of
conductance $(d^2 I/dV^2)$ with and without local heating. Counterexamples
are provided to show that the selection rule for important normal modes
based on the component of vibrations along the $z$-direction is inadequate.

We start with a brief introduction to the theories which lead to
the results of our calculations. We study the current-induced forces and
inelastic profiles in the framework of density functional theory (DFT)
in scattering approaches~\cite{Lang}. The wavefunctions
$\Psi_{E K_{\parallel}}^{\alpha}(\textbf{r})$ are calculated by solving
the Lippmann-Schwinger equation iteratively until
self-consistency is obtained.

The force $\bf{F}$, acting  on a given atom at position $\bf{R}$ exerted by
the nonequilibrium current, is given by the Hellmann-Feynman type of
theorem which has been developed in Ref.~\cite{Diventra}.
\begin{eqnarray}
\bf{F}=\sum_{i}\langle\psi_{i}|\frac{\partial H}{\partial \bf{R}}|\psi_{i}\rangle
+\lim_{\Delta\rightarrow 0}\int_{\sigma}dE\langle\psi_{\Delta}|\frac{\partial H}
{\partial \bf{R}}|\psi_{\Delta}\rangle,
\label{eq:force}
\end{eqnarray}
where the first term on the right hand side of Eq.~\ref{eq:force} is the Hellmann-Feynman
contribution to the force due to localized electronic states $|\psi_i\rangle$.
The second term is the contribution to the force from the continuum
states $|\psi_{\Delta}\rangle$.

The inelastic current and local heating are calculated using the
first-order perturbation theory based on the second-quantized formalism.
By applying the field operator with the wavefunctions obtained in DFT,
the many-body Hamiltonian of the system is $H=H_{el}+H_{vib}+H_{el-vib}$,
where $H_{el}$ is the electronic part under adiabatic approximations;
$H_{vib}$ is the ionic part of the Hamiltonian, which can be casted into a set
of independent simple harmonic oscillators via normal coordinates; and
$H_{el-vib}$ represents the electron-vibration interactions and has the form
\begin{eqnarray}
H_{el-vib} &=&\sum_{\alpha,\beta,E_{1},E_{2},j}\left(  \sum_{i,\mu}%
\sqrt{\frac{\hbar}{2M_{i}\omega_{j}}}A_{i\mu,j}J_{E_{1},E_{2}}^{i\mu
,\alpha\beta}\right) \nonumber\\
&&\times a_{E_{1}}^{\alpha\dag}a_{E_{2}}^{\beta}(b_{j}+b_{j}^{\dag
}),\label{elph}%
\end{eqnarray}
where $\alpha$ and $\beta$ refers to either the left (L) or the right (R) electrodes;
$M_i$ is the mass of the $i$-th atom; $A_{i\mu,j}$ is a canonical transformation
between normal and cartesian coordinates satisfying
${\displaystyle\sum\nolimits_{i,\mu}}A_{i\mu,j}A_{i\mu,j^{\prime}}=\delta_{j,j^{\prime}}$;
$\omega_{j}$ are the normal mode frequencies; $b_{j}$ is the annihilation operator
for phonons corresponding to the $j$-th normal mode; and $a^{L(R)}$is the annihilation
operator for electrons. The coupling constant $J_{E_{1},E_{2}}^{i\mu,\alpha\beta}$
between electrons and the vibration of the $i$-th atom in $\mu$-th $(\mu=x, y, z)$
component, i.e.,
\begin{equation}
J_{E_{1},E_{2}}^{i\mu,\alpha\beta}=\int d\mathbf{r}\int d\mathbf{K}%
_{\parallel}[\Psi_{E_{1}\mathbf{K}_{\parallel}}^{\alpha}(\mathbf{r})]^{\ast
}[\partial_{\mu}V^{ps}(\mathbf{r},\mathbf{R}_{i})\Psi_{E_{2}\mathbf{K}%
_{\parallel}}^{\beta}(\mathbf{r})],\label{couplingJ}%
\end{equation}
where $V^{ps}(\mathbf{r},\mathbf{R}_{i})$ is the pseudopotential representing
the interaction between the electron at $\mathbf{r}$ and the $i$-th ion
at $\mathbf{R}_{i}$.

The current in the presence of electron-vibration scattering within
the first-order correction is,~\cite{Hsu},
\begin{eqnarray}
I=\frac{2e}{h}\int dE
[(f_{E}^{R}-f_{E}^{L})-(\tilde{B}^{R}-\tilde{B}^{L})]\tau(E),\label{eq:8}
\end{eqnarray}
where $f_{E}^{L(R)}=1/[\exp\bigl((E-\mu_{L(R)})/(k_{B}T_{L(R)})\bigl)+1]$ is
the Fermi-Dirac distribution function for the left(right) electrode,
$k_{B}$ is the Boltzmann constant, $\mu_{L(R)}$ the chemical
potential in left(right) electrode, $T_{L(R)}$ is the temperature
in the left(right) electrode, and $\tau(E)=\frac{\pi\hbar^{2}}{mi} \int
d\textbf{R}\int d \textbf{K}_{\parallel}(\Psi^{R*}_{E}\nabla \Psi^{R}_{E}-\nabla \Psi^{R*}_{E} \Psi^{R}_{E})$
is the transmission function of electrons with the energy $E$ from the
right electrode. In the absence of electron-vibration interactions,
Eq.~(\ref{eq:8}) returns to the Landauer-B\"{u}ttiker current formula.
The normalized parameters $\tilde{B}^{\alpha}(\alpha=L,R)$ from the inelastic scattering
processes are obtained by the following equations:
\begin{eqnarray}
\tilde{B}^{\alpha}&=&\sum_{j\nu}[\langle|B_{j\nu,k}^{\beta\alpha}|^{2}\rangle
f_{E}^{\alpha}(1-f_{E\pm \hbar \omega_{j\nu}}^{\beta}),
\label{eq:balpha}
\end{eqnarray}
where $\{\alpha,\beta\}=\{L,R\}$ and $\alpha\neq\beta$. The parameter
$B_{j\nu,1(2)}^{\beta\alpha}$ in Eq.~(\ref{eq:balpha}) is,
\begin{eqnarray}
B_{j\nu,1(2)}^{\beta\alpha}&=&i\pi
\sum_{i\mu}\sqrt{\frac{\hbar}{2\omega_{j\nu}}}A_{i\mu,j\nu}J_{E\pm
\hbar\omega_{j\nu} ,E}^{i\mu,\beta\alpha}D^{\beta}_{E\pm \hbar
\omega_{j\nu}}\nonumber\\&&\times\sqrt{\delta+n_{j\nu}},
\end{eqnarray}
where $\delta=0(1)$ represents the process of
phonon absorption(emission), and $n_{j\nu}=1/[\exp(\hbar\omega_{j\nu}/k_{B}T_{w})-1]$ denotes
the ensemble averages of phonon states where $T_{w}$ is the local temperature.
$\langle B_{j\nu,k}^{\beta\alpha}\rangle$ refers to the ensemble average
over phonon states.

To incorporate local heating in our calculations, we compute the total thermal
power generated in the junction $P$ via electron-vibration interactions using~\cite{Hsu},
\begin{eqnarray}
P=\sum_{\j\in vib}\sum_{\alpha,\beta=\{L,R\}}(W_{j}^{\alpha\beta,2}-W_{j}^{\alpha\beta,1}),\label{eq:work}
\end{eqnarray}
where $W_{j}^{\alpha\beta,2(1)}$ is the power calculated up to the first order from
the Fermi golden rule, corresponding to relaxation (excitation)
of the vibrational modes~\cite{Chen}.

%We first perform vibrational analysis using \textit{Gaussian} 03.
\begin{figure}
\includegraphics[width=8.0cm]{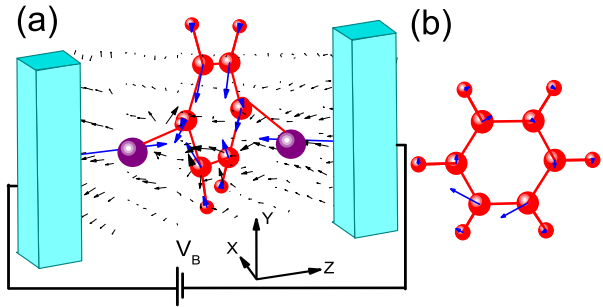}
\caption{
%\textbf{$\mid$ Current-induced forces under the curved flow of electron current streamline.}
(Color online).
(a) 3D force vector plot (blue line) of the benzene molecule (red) connected to
platinum atoms (purple) with $V_{B}=0.1$~V on top of the current density vector plot (black line).
The magnitude of the current density varies in orders of magnitude. The graph is
plotted in log scale to help the visualization of the smaller
current density. The separation of Pt-electrodes (jellium modeled, r$_{s}\approx3$)
is fixed at $9.626$~a.u.
(b) Projection of the force vectors on the plane of the benzene molecule.
The projected net torque around the center of benzene ring
is $(0.0013, -1.637\times10^{-7}, -0.0016)$~nN-{\AA}. }
\label{fig:force}
\end{figure}

We consider successively the effects of the curved flow of current streamline
on the current-induced forces and inelastic profiles in a highly asymmetric
single-molecule junction. First, we compare the current density
at $0.1$~V with the forces on each
atom in the Pt/benzene junction, where the current-induced forces are calculated by
subtracting the total forces at $0.1$~V and $0$~V, as shown in Fig.~\ref{fig:force}(a).
We observe that the benzene is subjected to compressional forces in both the horizontal
and vertical directions. This feature is consistent with the current-induced
forces for the conventional Au/benzene junction, where the benzene molecule lies on
the same plane with both electrodes and exhibits mirror symmetry~\cite{Diventracontact}.

In the Pt/benzene junction, the horizontal compressional force on the two outermost
platinum atoms is significantly stronger than the vertical force because the
current flows mostly through these two atoms. Due to the tilted benzene
ring with respect to the $y-z$ plane, the net force is considerably
stronger in the negative $y$ and negative $z$ directions.
The curved current streamline flow around the molecule produces
a net torque around the center of the benzene ring. The net torque
[around (0.010, -0.077, 0.005)~nN-{\AA}] is stronger in the negative $y$-direction.
This means that the force induced by the curved current stream line flow
could lead to a clockwise rotation in the $x-z$ plane.
We also project the force vectors onto the plane of the benzene ring, as
shown in Fig.~\ref{fig:force}(b).
We observe that the benzene ring experiences a net torque which
tends to rotate the molecule clockwisely. Thus, the highly asymmetric
single-molecule junction, such as the Pt/benzene junction, offers a
practical system in the exploration of the possibility that current
could be used to drive atomic-scale motors.

\begin{figure}[bp]
\includegraphics[width=8.0cm]{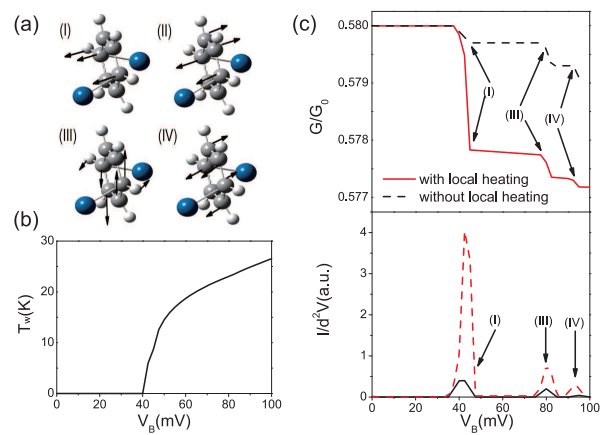}
\caption{
(Color online).
%\textbf{$\mid$ Schematic of normal modes and the inelastic profiles.}
(a) Four vibrational modes corresponding to energies (I)42~eV,
(II) 64~eV, (III) 80~eV, and (IV) 93~eV.
(b) Local temperature $T_w$ as a function
of the applied bias $V_{B}$ at zero electrode temperature.
(c) Conductance and the
derivative of conductance as a function of the applied bias $V_{B}$ with local heating (
red and dashed) and without local heating(black and solid).
Temperatures of electrodes are set to zero.}
\label{fig:IETS}
\end{figure}

Second, we show that the current streamline flow is crucial to explain the
selection rule for normal modes shown in the IETS. Without looking into the
details of current density, the selection only based on the current
could lead to misjudgement with regard to the importance of normal
modes for IETS in the Pt/benzene junction and other highly
asymmetric molecular junctions.

In Fig.~\ref{fig:IETS}(a), four modes denoted as (I), (II), (III) and (IV)
out of $42$ vibrational modes with normal mode energies around $42$~eV,
$64$~eV, $80$~eV, and $93$~eV, respectively, are shown. Mode~(I) is
acoustic-like and contributes to IETS at V$_{B}=42$~mV,
which has been experimentally observed~\cite{Kiguchi}. This mode corresponds
to a vibration of the benzene molecule as a whole, with a large component of
motion along the current. In other words, the center of mass of the benzene
molecule moves along the $z$-direction.
%It is interesting to note that this mode
%is only visible when including the platinum atoms from the first layer of the electrode
%in our calculation.
Mode~(II) is a longitudinal mode where the middle part vibrates
in the direction opposite to those of the top and the bottom parts.
Mode~(III) is a transverse mode where the left side of the carbon ring
vibrates downwards while the right-side vibrates upwards. This mode has also
been observed in the IETS but has not been discussed in Ref.~\cite{Kiguchi}.
Mode~(IV) is another longitudinal mode. However, only the second-nearest
neighbor vibrates in the same direction. The center of mass of the benzene
molecule remains fixed.

To include the effects of local heating, we obtain the rate of thermal energy
dissipated to the bulk electrodes via phonon-phonon interactions using
the weak-link model~\cite{Geller}. The stiffness of the benzene molecule is
estimated using the total energy calculation, which
gives $K=6.67302\times 10^{-4}$~eV$_0/a_0^2$, where $a_0$ is the Bohr radius.
The effective local temperature $T_{w}$ is achieved when heat generation in
the nanostructure and heat dissipation into the bulk electrodes are balanced~\cite{Chen}.
Fig.~\ref{fig:IETS}(b) shows the effective local temperature $T_w$ vs. V$_{B}$ in
a Pt/benzene junction, where the temperatures of bulk Pt electrodes are set to $0$~K.
We observe that the local temperature increases rapidly when the
bias is larger than V$_B$$=42$~mV, where the electrons have enough energy to excite
mode~(I), the lowest energy mode vibrating along the $z$-direction.
%This is consistent with the results obtained in a 4-Al atomic junctions where the local heating is stronger in the low temperature regime~\cite{Hsu}.

Subsequently, we calculate the inelastic profile of conductance ($dI/dV$)
[upper panel in Fig.~\ref{fig:IETS}(c)] and the absolute value of the differential
conductance ($d^2 I/dV^2$) [lower panel in Fig.~\ref{fig:IETS}(c)], with and without
the presence of local heating, as a function of bias voltage. The zero-bias
conductance (approximately $0.58$~G$_{0}$) is in agreement with the
large conductance observed in the experiments~\cite{Kiguchi}. This is
in sharp contrast to the small conductance which has been
found in the conventional junction, where the benzene molecule is
connected to the gold electrodes via the sulfur atoms.
When local heating is included, we observe that the inelastic features
are enhanced due to increased number of local phonons.
Figure~\ref{fig:IETS}(c) shows that large jumps in IETS correspond to three
normal modes [Modes~(I), (III), and (IV)]. Modes~(I) and (IV) clearly
show a large component of vibration along the direction
of electron transport (i.e., roughly along the line connecting two Pt atoms),
as shown in Fig.~\ref{fig:IETS}(a). The large contributions of these
two modes can be explained by the selection rule based on the current:
the strength of inelastic effects are related to the component of vibration
along the $z$-direction~\cite{chenH2}. However, Mode~(III) is an exception
to the selection rule based on the current. Fig.~\ref{fig:IETS}(a) indicates
that Mode~(III) is a transverse mode. The motion is mostly perpendicular
to the line connecting the two Pt atoms, and thus, its contribution to
the inelastic effects is expected to be minimal. The
large contribution of Mode~(III) to the IETS has posed a perplexing problem
regarding the applicability of the selection rule based on the current.

\begin{figure}
\includegraphics[width=8.0cm]{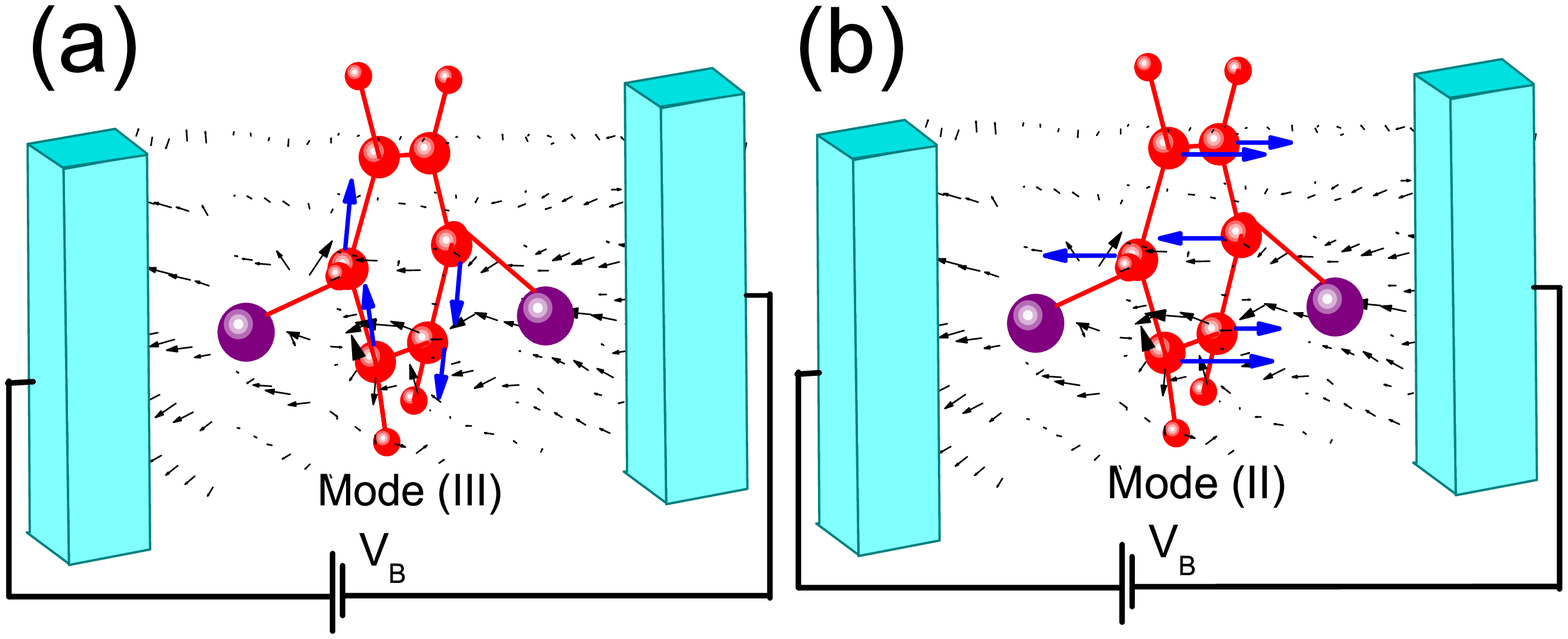}
\caption{
(Color online).
%\textbf{$\mid$ Examples of misjudgement in the importance of normal modes in the inelastic profiles.}
(a) Schematic of Mode~(III) and (b) Mode~(II)
with vibrational direction
(blue line) on top of the current density vector plot (black line)
of the benzene molecule (red) connected to the platinum atoms (purple)
sandwiched between two Pt electrodes with a bias voltage
of $V_{B}=0.1$~V.}
\label{fig:currentdensity}
\end{figure}

To settle this problem, we visualize the current density and compare
the electron current streamline flow with the motion of normal modes.
It should be noted that the center of the mass (CM) of the benzene
molecule does not lie in the line connecting the two platinum atoms.
Under such highly asymmetric configuration, the current tends to
flow through the bottom half of the benzene molecule, as shown
in Fig.~\ref{fig:currentdensity}.
For Mode~(III), the carbon ring near the left electrode vibrates
upward while the carbon-ring near the right electrode vibrates downward.
Fig.~\ref{fig:currentdensity}(a) clearly shows that the carbon-ring
vibrates along the curved electron current streamline flow.
This justifies the importance of Mode~(III) to the IETS.
We illustrate another example which shows the importance of
the current density on the selection rule for inelastic effects.
Figure~\ref{fig:IETS}(c) displays that Mode~(II) has a large component
of motion along the $z$-direction, but the corresponding inelastic
profile does not occur, as shown in Fig.~\ref{fig:IETS}(c).
Figure~\ref{fig:currentdensity}(b) shows that the motion
of Mode~(II) is mostly transverse to the curved current
streamline flow. Thus, this mode is unimportant to the IETS.

In conclusion, we have shown that the current-induced forces and
inelastic profiles are highly relevant to the details of current
density, especially in highly asymmetric molecular junctions.
The asymmetric current streamline flows may cause net torque,
which tends to rotate the molecule, in a manner similar to
a stream of water rotates a waterwheel. Thus, the highly asymmetric
single-molecule junctions offer practical systems to explore
the possibility of an atomic-scale motor driven by the non-equilibrium
electron transport. Current density is also important in the explanation
of the mode selection rule for inelastic effects. In the absence
of detailed information on the current density, we enumerate
examples to show that the importance of normal modes on the inelastic
effects could be misjudged.

The authors are grateful to Ministry of Education Aim at Top University
Plan(MOE ATU), National Center for Theoretical Sciences(South), and
National Science Council (Taiwan) for support (Grants NSC97-2112-M-009-011-MY3).
We thank National Center for High-performance Computing
for computer time and facilities.

\end{document}